\documentclass[]{elsarticle}
\pdfoutput=1
\usepackage[latin2]{inputenc}
\usepackage[T1]{fontenc}
\usepackage[]{amsmath}
\usepackage{xspace}
\usepackage{mathdots}
\def\PP{\textsc{P}\xspace}

\def\exptime{\textsc{ExpTime}\xspace}
\def\aexptime{\textsc{AExpTime}\xspace}
\def\nexptime{\textsc{NExpTime}\xspace}

\def\pspace{\textsc{PSpace}\xspace}
\def\expspace{\textsc{ExpSpace}\xspace}

\def\exwotime{\textsc{DExpWOTime}\xspace}
\def\aexwotime{\textsc{AExpWOTime}\xspace}
\def\nexwotime{\textsc{NExpWOTime}\xspace}
\def\exwospace{\textsc{DExpWOSpace}\xspace}

\def\nexwospace{\textsc{NExpWOSpace}\xspace}
\def\wopspace{\WO--\pspace}
\def\dwopspace{\textsc D\WO--\pspace}
\def\ktm#1{$k$--#1}
\def\WO{\textsc{WO}\xspace}
\def\Aa{\ensuremath{\mathcal A}\xspace}

\def\O#1{O\left(#1\right)}
\def\O#1{O(#1)}

\def\kexp#1#2{ \overbrace{2^{2^{\iddots^{n^#2}}}}^{#1+2}}
\newtheorem{thm}{Theorem}
\newtheorem{lem}[thm]{Lemma}
\newdefinition{rmk}{Remark}
\newproof{pf}{Proof}
\newproof{pot}{Proof of Theorem \ref{thm4} - continuation}
\begin{document}
\title{On multiply-exponential write-once Turing machines}
\tnotetext[t1]{This work is supported by Polish NCN 
grant~DEC-2012/07/B/ST6/01532.} 
\author[mim]{Maciej Zielenkiewicz\corref{corr1}}
\ead{maciej.zielenkiewicz@mimuw.edu.pl}
\author[mim]{Aleksy Schubert\corref{corr1}}
\ead{alx@mimuw.edu.pl}
\author[mim]{Jacek Chrząszcz\corref{corr1}}
\ead{chrzaszcz@mimuw.edu.pl}
\address[mim]{Faculty of Mathematics, Informatics and Mechanics, University of Warsaw, Poland}
\cortext[corr1]{Corresponding authors.}
\begin{abstract}
  In this work we analyze the multiply-exponential complexity classes for write-once Turing machines, i.e. machines that can write to a given tape cell at most once.  We show that \ktm\exwospace $=$ \ktm\exwotime $=$ \ktm\exptime and the nondeterministic counterpart.
For alternating machines we show that \ktm\aexwotime $=$ \ktm\aexptime $=k$-$1$--\expspace.
\end{abstract}

\maketitle
\section{Introduction}
The idea of write-once machines was first studied by Hao Wang \cite{HaoWang} in 1957. Those machines are analogous to the Turing machines, with the exception that writing is possible only to blank spaces; this model fits many types of storage which are used both today and historically, for example punch cards and tapes and recordable optical discs (CD-R, DVD-R etc.). Those machines are equivalent to Turing machines with respect to the languages they accept. In this paper we characterize some complexity classes of write-once machines and show what are the corresponding complexity classes for the usual Turing machines. Additionally, we consider a further restriction of write-once machines, that are allowed to write only at the end of the tape, and show that this class of machines is not universal.

\section{Known results}
In 1960 Lee \cite{Lee} has shown many useful conversions of programs for usual Turing machines to programs for write-only machines with their complexities. 
Subsequently Rivest and Shamir \cite{Rivest} have shown a coding scheme which allows for efficient simulation of updates; one of their most important results is that a value can be stored in a way which permits $t$ updates at the expense of increasing the storage size asymptotically $t/\log t$ times. We will be using this result extensively through the paper.  The coding schemes for efficient updates are still an active area of research (for example see \cite{Shpilka}), with new applications in efficient usage of flash memory, as the erase operation is much more expensive (i.e. is slower and causes degradation) than a simple write.

An analysis of complexity of some algorithms was done by Irani et al (\cite{Irani}), who show that an algorithm with a given complexity using $k$ space in the usual computing model can be rewritten to the write-only (deterministic) RAM machine model (with a constant amount of additional multiple-write memory) with an increase of running time by a factor of $\O{\log n/\log \log n}$ or by a factor of $\O{\log k}$. It is also shown that in general there is no simulation technique which uses less write-only space than the running time of the original algorithm. 
There are some important data structures which can be simulated as fast as the original algorithm, which include binary search trees and find-union data structures. Eventually in their paper it is shown that \wopspace$=$\PP, where $\WO$ denotes that the complexity class when using write-once memory. 


A slightly different model was presented by Vitter in \cite{Vitter}, where effects connected with usage of optical disks were taken into account. The model assumed a division of memory into blocks, where each write of a block would come with a need for additional memory for the block header, therefore efficient ways of aggregating writes need to be used. Their main result is that for block size $B$ and $n$ allocated regions on a disc the time needed for an I/O operation is $\O{\log n + B \log B}$.
\section{Basic definitions}\label{sec:definitions}

For simplicity we assume that the alphabet of the Turing machine is always just two symbols, and we call a cell blank ($0$) or marked ($1$). We can easily simulate a bigger alphabet by encoding a symbol in more than one cell.
In defining the write-once Turing machine we follow definitions set forth in the paper \cite{HaoWang} by Hao Wang, with a slight change that we will assume that the tape is infinite only in one direction. Again, the reduction to the machine with two-way infinite tape is not complicated, although it would have complicated the proofs unnecessarily --- we can pick a constant $C$ and have the machine mark every $C$th cell; we will call such a cell a length-marking cell. Then the machine can find the beginning of the tape by going to the last cell it has marked as length-marking, and repeat going $C$ cells to the left and checking if the cell is marked until we find an unmarked cell, which is the cell directly preceding the beginning cell of the tape.  
The definition of a write-once Turing machine is mostly the usual definition of the Turing machine, i.e.~it is a machine with one tape and one read/write head and a two symbol alphabet, but it is not allowed to write a blank to a cell which contains a non-blank symbol.
It is not an error to try to mark again the cell which is already marked, although it does not change the state of the tape.

The use of the Turing machine model is in contrast to the RAM machine model used in the analysis of \wopspace by Irani et al. (\cite{Irani}). 
We show how to prove their result in our model in Theorem \ref{pspace}. Our model naturally extends to non-deterministic and alternating versions, and we will be using letters ``D'' for deterministic, ``N'' for non-deterministic and ``A'' for alternating versions.

\section{Our reductions}
We show how some of the complexity classes for write-once machines relate to those for standard Turing machines. More specifically we prove that \linebreak\ktm\exptime $=$ \ktm\exwospace $=$ \ktm\exwotime. We also prove that the same equations hold if we replace all the classes with their nondeterministic counterparts.
Moreover, we show how \ktm\expspace relates to alternating write-once classes by showing that \ktm\aexwotime $=$ \ktm\aexptime $=k$-$1$--\expspace.

\begin{thm}
\dwopspace $=$ \PP.
\label{pspace}
\end{thm}
\begin{pf}
See proof of theorem~\ref{thm4} below, substituting 0 for $k$.
\end{pf}

\begin{thm}
\ktm\exptime $=$ \ktm\exwospace.
\label{thm4}
\end{thm}
\begin{pf}
First we show that \ktm\exptime $\subseteq$ \ktm\exwospace. Suppose we have a \ktm\exptime machine which has running time on input of length $n$ bounded by the function $f(n)=\kexp k t$. Then it can write a tape of length at most $f(n)$ updating each location at most $f(n)$ times. To store a representation of a tape of this size permitting the expected number of updates we will need at most $f(n) \cdot f(n) = \kexp k t \cdot \kexp k t = \O{\kexp k {t+1}}$ memory, which is in \ktm\exwospace.

The other inequality is \ktm\exptime $\supseteq$ \ktm\exwospace, which is more tricky. Let \Aa be our \ktm\exwospace machine which makes at most $g(n)$ writes. 
We need to show how many steps are needed to simulate the execution between writes. We show a lemma for that:
\end{pf}

\begin{lem}
	A $k$-\exwospace machine $\Aa$ with space complexity $f(n) = \kexp k t$ started in a configuration in which it eventually stops makes at most $\O{f(n)\cdot s}$ steps between any two consecutive writes.
\end{lem}
\begin{pf}
	Suppose that $\Aa$ has $s$ states. Between two consecutive writes we can treat the machine as a two-way DFS automaton (the tape does not change). At most $f(n)$ of tape is used at any time. The automaton may visit a location on the tape at most $s$ times, because otherwise some location on tape is visited at least twice in the same state between two consecutive writes. The sequence of states between the two visits of the same location in the same state could be repeated any number of times, therefore the automaton would not stop~--- contradiction with assumption. Therefore there are at most $f(n)\cdot s$ steps between any two consecutive writes.
	\label{l1}
\end{pf}
\begin{lem}
	A $k$-\nexwospace machine $\Aa$ with space complexity $f(n) = \kexp k t$ started in a configuration in which it eventually stops has a run that makes at most $\O{f(n)\cdot s}$ steps between any two consecutive writes.
\end{lem}
\begin{pf}
The proof is very similar to this in Lemma \ref{l1}, but we show how to shorten the runs instead of showing nontermination. More specifically, we can simulate the operations of the non-deterministic two-way finite automaton between two consecutive writes. In case the automaton is twice in the same state in a given location on the tape we can just skip the computation steps between the two visits. This manipulation performed iteratively results in a run that visits each location in each state at most once. The number of steps is therefore $f(n)\cdot s$. 
	\label{l1N}
\end{pf}

\begin{pot}
Our simulation is in \ktm\exptime, as the time needed for simulation can be computed as 
\[ (\textrm{\# of writes})\cdot(\textrm{\# of steps between two writes}) = g(n) \cdot O(g(n)) = O((g(n))^2). \]
\end{pot}

\begin{thm}
\ktm\nexptime $=$ \ktm\nexwospace.
\label{thm4n}
\end{thm}
\begin{pf}
The proof is the same as the proof of the Theorem \ref{thm4}, except that we have to use Lemma \ref{l1N} instead of Lemma \ref{l1}.
\end{pf}

\begin{thm}
\ktm\exwotime $=$ \ktm\exptime.
\end{thm}
\begin{pf}
Notice that \ktm\exwotime $\subseteq$ \ktm\exptime is trivial, since the only difference between write-once machine and usual Turing machine is a \emph{restriction} on the eligible transitions. \\
To prove \ktm\exwotime $\supseteq$ \ktm\exptime we need to show how to simulate a \ktm\exptime machine with a \ktm\exwotime one. Suppose that the \ktm\exptime machine has time complexity of $f(n)=\kexp k n$. The only difficulty with simulating the writes is the restriction of non-overwriting memory, but we can copy the contents of the entire tape on each write that changes a marked cell to blank. The cost of simulation of a single write is $f(n)^2$, and reads can be performed without any additional steps. Therefore
the running time will be $O((f(n))^3)$, which does not drive us outside the \ktm\exptime complexity class.
\end{pf}

\begin{thm}
\ktm\nexwotime $=$ \ktm\nexptime and 
\ktm\aexwotime $=$ \ktm\aexptime ($=(k$-$1)$-\expspace).
\end{thm}
\begin{pf}
The above construction remains unchanged if we consider nondeterministic/alternating automata. In case of the alternating automata, the alternating states can be rewritten to use ``fresh'' memory.
\end{pf}

\subsection{A note on automata writing at the end of tape.}
A natural next step of our model simplification is to consider automata which write only at the end of the tape. We will show that such automata are not universal.
Suppose we have an alphabet with a blank symbol and at least two non-blank symbols. We can introduce a restriction which does not allow blank symbol in the portion of the tape which was already written, i.e. no blanks between non-blank symbol. We will show how to simulate an automaton with this restriction while not keeping a full copy of its memory.

First we convert our machine so that it always goes from left to right and then to the beginning of the tape and so on. Then we show a pumping lemma operating on the automaton's memory.

\begin{lem}
	For a given \WO-automaton \Aa with $k$ states which writes only at the end of the tape, suppose $s$ is a state in which \Aa writes to the tape. Then there exists an automaton $\Aa_s$ which simulates the run of \Aa from $s$ to the next writing state, which moves to the beginning of the tape and then only reads the tape from left to right.
\label{l2a}
\end{lem}
\begin{pf}
When running from $s$ to the next writing state, \Aa does no writes, so can be seen as a simple automaton with only input and not output (the symbol to be written can be stored in the state). Results of Vardi \cite{Vardi} show that such an automaton can be converted to one-way automaton at the expense of increasing the number of states to $\O{\exp k}$.
\end{pf}

\begin{lem}[Pumping lemma]
	For write-only automaton \Aa which writes only at the end and has $k$ states, we can pump any word of length at least $p(k)=\O{2^{\exp(k)^k}}$ with respect to runs between writes with a single pumping scheme for all states.
	\label{l3}
\label{pumpl}
\end{lem}
\begin{pf}
We have at most $k$ writing states. Using Lemma~\ref{l2a} we build automata for all of these states. Then we pump the product automaton of all those automata, which has $K=(\exp(k^2))^k$ states, so words longer than $O(2^K)$ can be pumped down.
\end{pf}
We use our pumping lemma (Lemma \ref{l3}) to shorten the runs between writes. The pumping length depends only on the number of states, so the maximum run length which we would need to store is limited by a constant; however we need some operations to do the pumping. We will run the automaton for $2\cdot p(n)$ steps from the beginning and for $p(n)$ steps backwards from the end (this needs nondeterminism, but we can simulate it with deterministic automaton at an exponential cost in term of the number of states) and find the state which occurs at least twice (positions $i_1$, $i_2$) in the normal run and at least once ($i_3$) in the backwards run. Then we can conclude that the tape could have been pumped down between positions $i_1$ and $i_3$, so we do not need to simulate the automaton any further in the reading run and we can continue from the writing state. If there is more than one writing state we simply have to have a bigger set of states in the beginning of the backwards simulation.

Then we do not need to store the whole tape contents of an automaton to simulate its run, as we can pump the tape down, and we know that the portion which was pumped down will not change. This means that a machine of this class can be simulated with an amount of memory which depends only on the number of states plus the size of the input, so the acceptance problem for this class of machines can be decided in \pspace.

\section{Conclusions}
We have analyzed the multiply-exponential complexity classes for write-once Turing machines and shown basically that, as far as such big complexities are concerned, write-once Turing machines perform as good (or as bad) as the regular Turing machines. The only difference is that corresponding space and time classes are equal for write-once machines, whereas this remains unknown for regular Turing machines. If further restriction on write-once machines is applied, namely that writing is allowed only at the end of the tape, the resulting class of machines is not universal.

\bibliography{write-once}
\bibliographystyle{elsarticle-num}
\end{document}